\documentclass[aps,showpacs,twocolumn]{revtex4}
\usepackage{epsfig}

\begin{document}

\title{Study of $p\bar{\Lambda}$ and $p\bar{\Sigma}$ systems in constituent quark models}

\author{Hongxia Huang$^1$, Jialun Ping$^1$ and Fan Wang$^2$}

\affiliation{$^1$Department of Physics, Nanjing Normal University, Nanjing
210097, P.R. China \\
$^2$Department of Physics, Nanjing University, Nanjing 210093, P.R. China}

\begin{abstract}
The $p\bar{\Lambda}$ systems with $J=0$ and $J=1$ are dynamically
investigated within the framework of two constituent quark models:
the chiral quark model and the quark delocalization color
screening model. The model parameters are taken from our
previous work, which gave a good description of the
proton-antiproton $S$-wave elastic scattering cross section
experimental data. The $p\bar{\Lambda}$ elastic
scattering processes with coupling to $p\bar{\Sigma}$ state
are studied. The results show that, there is no
$s$-wave bound state as indicated by an enhancement near the
threshold of $p\bar{\Lambda}$ in $J/\psi$ decay. However,
a $IJ = \frac{1}{2}0$ $p\bar{\Sigma}$ resonance state is given in
the quark delocalization color screening model.
\end{abstract}


\maketitle

\section{INTRODUCTION}

The recent observation of a near-threshold narrow enhancement in
the $p\bar{p}$ invariant mass spectrum from radiative decay
$J/\Psi \rightarrow \gamma p\bar{p}$ by the BES Collaboration
\cite{BES} has renewed interest in the $N\bar{N}$ interaction and
its possible baryonium bound states. In fact the enhancement in
the $p\bar{p}$ invariant mass distribution near the threshold
has been observed by the Belle Collaboration in the decays $B^+\rightarrow
K^+p\bar{p}$ and $B^0\rightarrow D^0p\bar{p}$~\cite{Belle}. Besides,
The Belle Collaboration also observed a near-threshold enhancement in the
$p\bar{\Lambda}$ invariant-mass spectrum in $B \rightarrow
p\bar{\Lambda}\pi$ decays~\cite{Belle2}. Then, the same enhancement
near the $p\bar{\Lambda}$ mass threshold was observed in the combined
$p\bar{\Lambda}$ and
$\bar{p}\Lambda$ invariant-mass spectrum from $J/\Psi \rightarrow
pK^{-}\bar{\Lambda}+c.c.$ decays by
BES Collaboration and it can be fitted with an $s$-wave
Breit-Wigner resonance with a mass $m = 2075 \pm 12(stat) \pm
5(syst)$ MeV and a width of $\Gamma = 90 \pm 35(stat) \pm 9(syst)
$ MeV or with a $P$-wave Breit-Wigner resonance~\cite{BES2}.
However, there is no significant signal in
$B^-\rightarrow J/\psi \Lambda \bar{p}$~\cite{Belle3}. It is,
therefore, of special interest to search for possible resonant
structures in other baryon-antibaryon states. In Ref.\cite{nonet},
Fermi-Yang-Sakata-like scheme was used to classify the possible
baryon-antibaryon SU(3) nonets. A systematic search of baryon-antibaryon
states in color-magnetic interaction model was also performed and
several interesting states were proposed~\cite{Ding}.

Quantum chromodynamics (QCD) is widely accepted as the fundamental
theory of the strong interaction, so it is naturally to expect to
understand hadron-hadron interaction from QCD. However, the direct
use of QCD on the hadron-hadron interaction is too difficult
because of the non-perturbative complications of QCD in the low
energy region. Recently, lattice QCD, Dyson-Schwinger approach and other
non-perturbative methods have made impressive progresses~\cite{Lattice,DSE},
but it is still far from satisfactory.
QCD-inspired quark models are the main tool for detailed studies
of the hadron-hadron interaction and multiquark systems at the moment.
The commonly used quark model is the constituent quark model,
where the complicated interactions between current quarks are
approximately transformed into dynamic properties of
quasi-particles (constituent quarks) and the residual interactions
between these quasi-particles. The multi-gluon effect and other
nonperturbative properties of QCD are attributed to the phenomenological
confinement potential between constituent quarks. The residue interactions
include effective one gluon exchange and one-Goldstone-boson exchange~\cite{manohar}.
The constituent quark model gives a good
description of properties of hadrons: meson ($q\bar{q}$) and
baryon ($q^{3}$), because of their unique color structures.
Applying to nucleon-nucleon scattering, a reasonable agreement
with experimental data is still possible after including the
$\sigma$-meson exchange for the chiral quark model
(ChQM)~\cite{chiralmodel1,chiralmodel2,chiralmodel3}, although
there is a controversy about its effect when taking $\sigma$ meson
as a $\pi\pi$
$S$-wave resonance~\cite{sigma}. Another constituent quark model
approach is the quark delocalization color screening model (QDCSM)
\cite{QDCSM1}, which has been developed with the aim of
understanding the well-known similarities between nuclear and
molecular forces despite the obvious energy and length scale
differences. In this model, two ingredients: quarks delocalization
and color screening are introduced to enlarge the Hilbert space
and to change the interaction between quarks resident in different
baryons and the delocalization parameter that appears is determined by the
dynamics of the interacting quark system. Thus the quark system can reach
its more favorable configuration through its own dynamics. The main
difference between the ChQM and the QDCSM is the mechanism of
intermediate-range attraction. The recent calculations showed that
both models can give a good description of the low-energy nucleon-nucleon
and hyperon-nucleon scattering~\cite{chenlz}, although they gave a little different
dibaryon resonance structures~\cite{wong,cm}. For
$N\bar{N}$ interaction, almost the same results are also obtained in both
models~\cite{pp,PRC73,pang,CTP43}, and there is no bound states as
indicated by a strong enhancement at threshold of $p\bar{p}$ in
$J/\Psi$ and $B$ radiative decay.
Therefore, extending the calculations to
$p\bar{\Lambda}$ study is an interesting practice.

In this paper, we study the
$p\bar{\Lambda}$ system by using both ChQM and QDCSM. It is quite
meaningful to investigate the difference of these two models in
baryon-antibaryon interaction. A brief description of these two
quark models of the baryon-antibaryon interaction is given in Section
2. The calculated results and discussions are given in Section 3.
Section 4 contains a brief summery.

\section{TWO QUARK MODELS}

\subsection{Chiral quark model}

The Salamanca version of ChQM is used in the present calculation.
The model details can be found in
Ref.\cite{chiralmodel1,PRC62,hypertriton}. To extend model from baryon-baryon systems to
baryon-antibaryon systems, the annihilation terms (gluon induced and
Goldstone boson induced), in addition to
scattering terms, have to be taken into account. The detailed description of
annihilation interaction has been given in Ref.\cite{PRC73,PRD26}.
The exchange interaction between quark and antiquark can be obtained by the
quark-antiquark symmetry. Here we only
write down the Hamiltonian for nucleon-antihyperon systems (the annihilation terms
take the same form as the ones of Ref.\cite{PRC73} because there is no $s\bar{s}$
annihilation term in the nucleon-antihyperon system),
\begin{eqnarray}
H &=& \sum_{i=1}^6 \left(m_i+\frac{p_i^2}{2m_i}\right)-T_{CM}
+\!\! \sum_{i<j=1}^6 \!\! V(\mathbf{r}_{ij})  \\
V(\mathbf{r}_{ij})& = &
V^{c}(\mathbf{r}_{ij})+V^{e}(\mathbf{r}_{ij})
 +V_{q\bar{q}}^{a}(\mathbf{r}_{ij}) \nonumber \\
V^{e}(\mathbf{r}_{ij}) & = & V_{qq(\bar{q})}^{Ge}(\mathbf{r}_{ij})
 +V_{qq(\bar{q})}^{\chi e}(\mathbf{r}_{ij})
+V^{s e}(\mathbf{r}_{ij}),  \nonumber \\
V_{q\bar{q}}^{a}(\mathbf{r}_{ij}) & = &
V_{q\bar{q}}^{Ga}(\mathbf{r}_{ij})
+ V_{q\bar{q}}^{\chi a}(\mathbf{r}_{ij}) \nonumber \\
V_{qq}^{c}(\mathbf{r}_{ij}) & = & -a_c \boldmath{\mbox{$\lambda$}}_{i}
\cdot
 \boldmath{\mbox{$\lambda$}}_{j}
 (r^2_{ij}+V_0) \nonumber \\
V_{q\bar{q}}^{c}(\mathbf{r}_{ij}) & = & a_c \boldmath{\mbox{$\lambda$}}_{i}
\cdot
 \boldmath{\mbox{$\lambda$}}^{*}_{j} (r^2_{ij}+V_0) \nonumber \\
V_{qq}^{Ge}(\mathbf{r}_{ij}) & = & \frac{1}{4}\alpha_s
\boldmath{\mbox{$\lambda$}}_{i} \cdot
 \boldmath{\mbox{$\lambda$}}_{j} \left[\frac{1}{r_{ij}}-\frac{\pi}{2}
 \right. \nonumber \\
 & &  \left. \left(\frac{1}{m^{2}_{i}}
 +\frac{1}{m^{2}_{j}}  +\frac{4 \boldmath{\mbox{$\sigma$}}_i\cdot
 \boldmath{\mbox{$\sigma$}}_j}{3m_{i}m_{j}}
  \right)\delta(\mathbf{r}_{ij})\right] \nonumber \\
V_{q\bar{q}}^{Ge}(\mathbf{r}_{ij}) & = & -\frac{1}{4}\alpha_s
\boldmath{\mbox{$\lambda$}}_{i} \cdot
 \boldmath{\mbox{$\lambda$}}^{*}_{j} \left[\frac{1}{r_{ij}}-\frac{\pi}{2}
 \right. \nonumber \\
 & & \left. \left(\frac{1}{m^{2}_{i}}
 +\frac{1}{m^{2}_{j}}+\frac{4\boldmath{\mbox{$\sigma$}}_i\cdot
 \boldmath{\mbox{$\sigma$}}_j}{3m_{i}m_{j}}
 \right) \delta(\mathbf{r}_{ij})\right]  \\
V_{qq}^{\chi e}(\mathbf{r}_{ij}) & = &
 v_{\pi}^e(\mathbf{r}_{ij}) \sum_{a=1}^3 f_i^a f_j^a
+v_{K}^e(\mathbf{r}_{ij}) \sum_{a=4}^7 f_i^a f_j^a
\nonumber \\
& & +v_{\eta}^e(\mathbf{r}_{ij}) (f^8_i f^8_j\cos \theta_P
 -f^0_i f^0_j\sin \theta_P) \nonumber \\
V_{q\bar{q}}^{\chi e}(\mathbf{r}_{ij}) & = &
 v_{\pi}^e(\mathbf{r}_{ij}) \sum_{a=1}^3 f_i^a f_j^{a*}
+v_{K}^e(\mathbf{r}_{ij}) \sum_{a=4}^7 f_i^a f_j^{a*}
\nonumber \\
& & +v_{\eta}^e(\mathbf{r}_{ij}) (f^8_i f^{8*}_j\cos \theta_P
 -f^0_i f^{0*}_j\sin \theta_P) \nonumber \\
v_{\chi}^e(\mathbf{r}_{ij}) & = &
\frac{g^2_{ch}}{4\pi}\frac{m^3_{\chi}}{12m_im_j}
\frac{\Lambda^{2}_{\chi}}{\Lambda^{2}_{\chi}-m_{\chi}^2}
\boldmath{\mbox{$\sigma$}}_i\cdot
 \boldmath{\mbox{$\sigma$}}_j \nonumber \\
& & \left[ Y(m_\chi r_{ij})-
\frac{\Lambda^{3}_{\chi}}{m_{\chi}^3}Y(\Lambda_{\chi} r_{ij})
\right]  ~~~ \chi=\pi,K,\eta. \nonumber \\
V^{s e}(\mathbf{r}_{ij}) & = & -\frac{g^2_{ch}}{4\pi}
\frac{\Lambda_{sca}^{2}}{\Lambda^{2}_{sca}-m_{sca}^2}m_{sca}
 \nonumber \\
 & & \left[ Y(m_{sca}
r_{ij})-\frac{\Lambda_{sca}}{m_{sca}}Y(\Lambda_{sca}
r_{ij}) \right],
 \nonumber \\
V_{q\bar{q}}^{Ga}(\mathbf{r}_{ij})& = & \frac{\pi}{6}\alpha'_{s}
 (\frac{16}{3}-\boldmath{\mbox{$\lambda$}}_{i} \cdot
 \boldmath{\mbox{$\lambda$}}^{*}_{j})
 (\frac{1}{3}+\frac{1}{2}\mathbf{f}_{i} \cdot \mathbf{f}^{*}_{j})
 \nonumber \\
 & & (3 +\boldmath{\mbox{$\sigma$}}_i\cdot
 \boldmath{\mbox{$\sigma$}}_j)
 \frac{\delta(\mathbf{r}_{ij})}{(m_i+m_j)^2}\nonumber \\
V_{q\bar{q}}^{\chi a}(\mathbf{r}_{ij}) & = &
c_{p}(\frac{1}{3}+\frac{1}{2}\boldmath{\mbox{$\lambda$}}_{i}
\cdot
 \boldmath{\mbox{$\lambda$}}^{*}_{j})(\frac{16}{9}-\frac{1}{3}\mathbf{f}_{i} \cdot
 \mathbf{f}^{*}_{j}) \nonumber \\
 & & (-\frac{1}{2}+\frac{1}{2}\boldmath{\mbox{$\sigma$}}_i\cdot
 \boldmath{\mbox{$\sigma$}}_j)\delta(\mathbf{r}_{ij}), \nonumber
\end{eqnarray}
Here, all symbols have their usual meanings. $Y(x)$ is the
standard Yukawa function. $Ge$ and $Ga$ ($\chi e$ and $\chi a$)
stand for one-gluon (Goldstone boson) exchange and annihilation
interactions, respectively. $V^{s e}$ is the effective scalar meson
exchange potential. When dealing with the strange system, the scalar
octet have to been considered. According to Ref.\cite{hypertriton}, the
effect of the scalar octet can be effectively taken into account
by s single scalar exchange potential $V^{s e}$ with different
parametrization for spin-singlet and spin-triplet channels (see Table 1 below).
According to QCD, the strong coupling constant $\alpha_s$ should be
scale dependent. In the constituent quark model, different strong
coupling constants: $\alpha_{s_{uu}},\alpha_{s_{us}}$ and $\alpha_{s_{ss}}$
($u,d$ quarks are taken as the same), are used for different interacting
quark pair: $uu,us$ and $ss$.

\subsection{Quark delocalization, color screening model}

\begin{table}[ht]
\caption{Parameters of the two quark models used. the masses of $\pi,K,\eta$
take their experimental values, $m_{\pi}=0.7$ fm$^{-1}$,
$m_{K}=2.51$ fm$^{-1}$, $m_{\eta}=2.77$ fm$^{-1}$.}
\centering
\begin{tabular}{lcc}\hline
 & {\rm ChQM} & ~~{\rm QDCSM}      \\
\hline
$m_{u,d}({\rm MeV})$        &  313    &  313     \\
$m_{s}({\rm MeV})$          &  573    &  573     \\
$b ({\rm fm})$              &  0.518  &  0.518  \\
$ a_c({\rm MeV\,fm}^{-2})$  &   48.59 & 58.03   \\
$ V_0({\rm fm}^{2})$        &   -1.2145 & -1.2883  \\
$\mu ({\rm fm}^{-2})$       &   --       &  0.5  \\
$\alpha_{s_{uu}}$           &  0.565  &  0.565  \\
$\alpha_{s_{us}}$           &  0.524  &  0.524  \\
$\alpha_{s_{ss}}$           &  0.451  &  0.451  \\
$\frac{g_{ch}^{2}}{4\pi}$   &  0.54  & 0.54   \\
$m_{sca}$ (fm$^{-1}$) (spin 0)  &  3.73    &  --     \\
$\Lambda_{sca} ({\rm fm}^{-1})$ (spin 0)  &  4.2    & --      \\
$m_{sca}$ (fm$^{-1}$) (spin 1)  &  4.12    &  --     \\
$\Lambda_{sca} ({\rm fm}^{-1})$ (spin 1)  &  5.2    & --      \\
$\Lambda_{\pi} ({\rm fm}^{-1})$   &  4.2    &  4.2     \\
$\Lambda_{K,\eta} ({\rm fm}^{-1})$   &  5.2    &  5.2     \\
$\theta_{P}$   &  $-15^{\circ}$    &   $-15^{\circ}$   \\
\hline
\end{tabular}
\end{table}

The model and its extension were discussed in detail in
Refs.\cite{QDCSM1,PRC65}. Its Hamiltonian has the same form as
Eq.(1), but with $V^{s e}=0$ and a different confinement
potential is used,
\begin{eqnarray}
V^{C}(\mathbf{r}_{ij})&=& -a_c \boldmath{\mbox{$\lambda$}}_{i}
\cdot
 \boldmath{\mbox{$\lambda$}}_{j} [f(r_{ij})+V_0], \nonumber
\\
 f(r_{ij}) & = &  \left\{ \begin{array}{ll}
 r_{ij}^2 & ~~~\mbox{if }i,j\mbox{ occur in the same} \\
 & \mbox{~~~baryon orbit}, \\
 \frac{1 - e^{-\mu r_{ij}^2} }{\mu} & ~~~
 \mbox{if }i,j\mbox{ occur in different} \\
  & \mbox{~~~baryon orbits},
 \end{array} \right.
\end{eqnarray}
$\mu$ is the color screening constant which to be
determined by fitting the deuteron mass.

The quark delocalization in QDCSM is realized by writing the
single particle orbital wave function of QDCSM as a linear
combination of left and right Gaussians, the single particle
orbital wave functions in the ordinary quark cluster models,
\begin{eqnarray}
\psi_{\alpha}(\mathbf{S}_i ,\epsilon) & = & \left(
\phi_{\alpha}(\mathbf{S}_i)
+ \epsilon \phi_{\alpha}(-\mathbf{S}_i)\right) /N(\epsilon), \nonumber \\
\psi_{\beta}(-\mathbf{S}_i ,\epsilon) & = &
\left(\phi_{\beta}(-\mathbf{S}_i)
+ \epsilon \phi_{\beta}(\mathbf{S}_i)\right) /N(\epsilon), \nonumber \\
N(\epsilon) & = & \sqrt{1+\epsilon^2+2\epsilon e^{-S_i^2/4b^2}}. \label{1q} \\
\phi_{\alpha}(\mathbf{S}_i) & = & \left( \frac{1}{\pi b^2}
\right)^{3/4}
   e^{-\frac{1}{2b^2} (\mathbf{r}_{\alpha} - \mathbf{S}_i/2)^2} \nonumber \\
\phi_{\beta}(-\mathbf{S}_i) & = & \left( \frac{1}{\pi b^2}
\right)^{3/4}
   e^{-\frac{1}{2b^2} (\mathbf{r}_{\beta} + \mathbf{S}_i/2)^2}. \nonumber
\end{eqnarray}
where $\mathbf{S}_i/2(-\mathbf{S}_i/2)$ is the reference center of
right (left) cluster.

Table 1 gives the model parameters used. the same values
of parameters: $b,~\alpha_{s_{uu}},~\alpha_{s_{us}},~\alpha_{s_{ss}},
~m_u,~m_s,~\cdots$, are used for both models. Thus, the two models
have the complete same contributions from one-gluon-exchange and
$\pi$, $K$, $\eta$
exchange. The only difference of the two models is coming from
intermediate-range part, A effective single scalar exchange
for chiral quark model, quark delocalization color screening for QDCSM.
The running property of the one-gluon coupling
constant is realized by the different values for $u,d,s$ quarks.
All the masses of the Goldstone bosons take the experimental values and
other parameters are determined by fitting baryons,
nucleon-nucleon interaction and deuteron properties except the
parameters $\alpha'_{s}$ and $c_{p}$ for the annihilation
interactions.
The one-boson annihilation coupling constant takes the value
$c_{p}=-0.2362~({\rm fm}^{2})$. The gluon annihilation coupling
constant $\alpha'_{s}$ is determined by fitting
the nucleon-antinucleon scattering cross sections~\cite{pp}.

\section{RESULTS AND DISCUSSIONS}

The $S$-wave $p\bar{\Lambda}$ systems with spin $J=0$ and $J=1$
are investigated in both the ChQM and the QDCSM. First the effective
potential between $p$ and $\bar{\Lambda}$ is calculated. The attractive
potential between two clusters is necessary to form a bound state.
The effective potential between two clusters is defined as
\begin{equation}
V_{eff}(s)  =  E(s)-E(\infty),
\end{equation}
where $s$ is the separation between the reference centers of two clusters
and $E(s)$ is total energy of the system,
\begin{equation}
E(s) = \frac{\left\langle \left[ \Psi_p \Psi_{\bar{\Lambda}}
\right]^{[222]J}_{W_cM_J} \right|
 H \left| \left[ \Psi_p \Psi_{\bar{\Lambda}} \right]^{[222]J}_{W_cM_J} \right\rangle}
 {\left\langle \left[ \Psi_p \Psi_{\bar{\Lambda}} \right]^{[222]J}_{W_cM_J} \right|
 \left. \left[ \Psi_p \Psi_{\bar{\Lambda}} \right]^{[222]J}_{W_cM_J} \right\rangle} .
\end{equation}
Fig. 1 shows the effective potentials for the
$p\bar{\Lambda}$ systems with $J=0$ and $J=1$. Clearly from Fig. 1,
we can see that for both $J=0$ and $J=1$ states, the effective potentials
are all attractive in both two models, so it is possible to form a bound state.
From Fig.1, we can also see that the potentials without
annihilation interactions are much more attractive than those with
the annihilation interactions, so the annihilation interactions provide
effective repulsion, which is consistent with the results of Ref.\cite{PRC73,PRD26}.

\begin{figure*}[ht]
\centering
\epsfxsize=5.5in \epsfbox{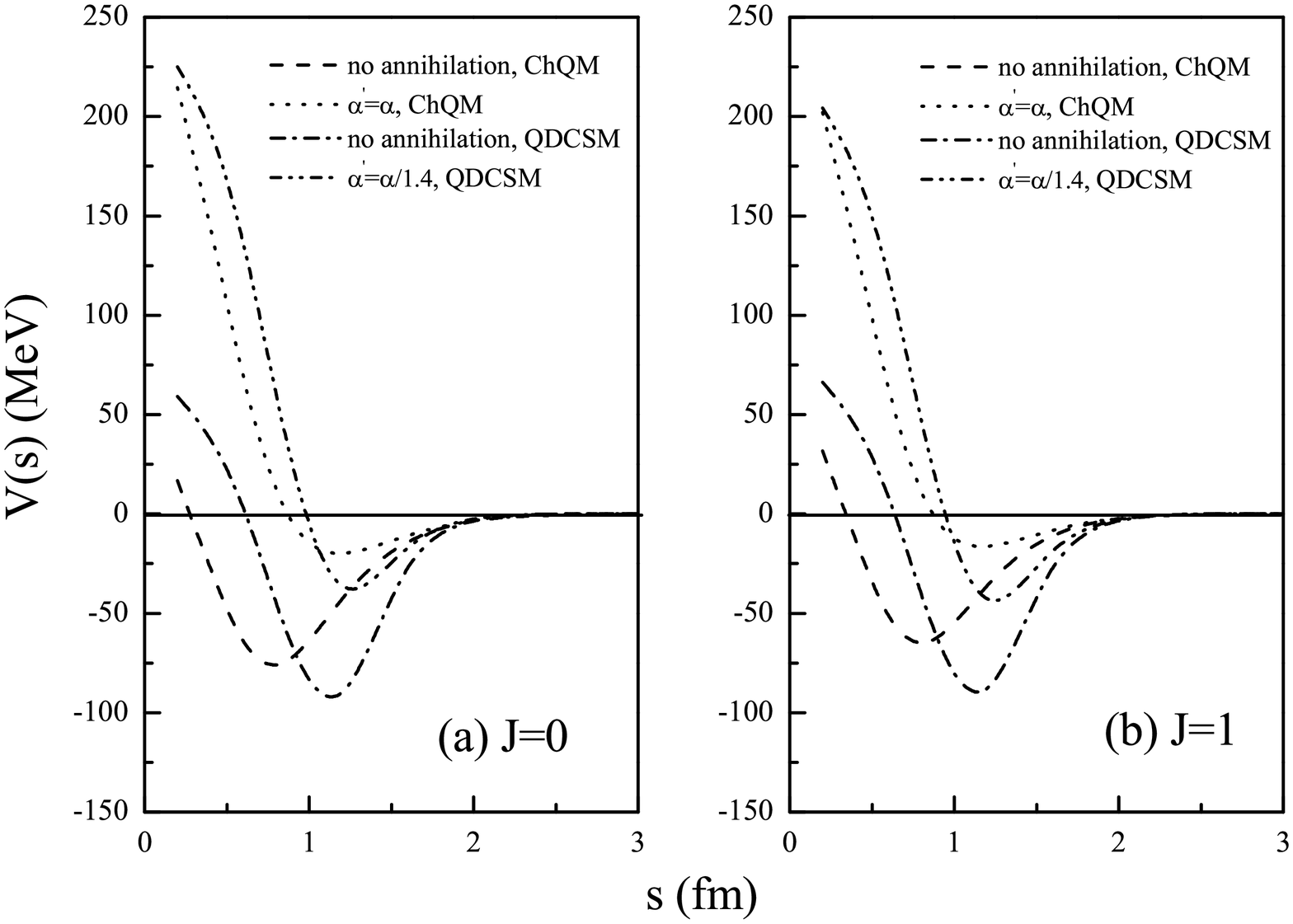}
\caption{The effective potential for an $S$-wave $p\bar{\Lambda}$
system in ChQM and QDCSM.}
\end{figure*}

Then, we do a dynamical calculation in both models by solving
bound-state RGM equation. We find that for both $J=0$ and $J=1$
states, there is no bound state in both two models if the annihilation
interaction terms are taken into account. However, the
effective potentials without annihilation interactions are deep enough to make
bound states for the $p\bar{\Lambda}$ systems. The calculated
binding energies are: $B_{p\bar{\Lambda}}=-14.1$ MeV for $J=0$ and
$B_{p\bar{\Lambda}}=-12.5$ MeV for $J=1$ in QDCSM;
$B_{p\bar{\Lambda}}=-14.7$ MeV for $J=0$ and $B_{p\bar{\Lambda}}=-13.2$ MeV
for $J=1$ in ChQM (Although the effective potentials in ChQM are a little
shallow, they have larger width, so almost the same binding energies
are obtained in two models). Here, the binding energy $B_{p\bar{\Lambda}}$
is defined as:
\begin{eqnarray}
B_{p\bar{\Lambda}}&=&
E_{p\bar{\Lambda}}-(M_{p}+M_{\bar{\Lambda}})
\end{eqnarray}

In order to search for $p\bar{\Lambda}$ bound state in a larger
space, a channel coupling calculation is performed. Here all
the possible color-singlet channels with strangeness 1 and spin 0,1
are taken into consideration. In the calculation, we find the effect of
channel-coupling for $p\bar{\Lambda}$ is so small that it can be
neglected safely. However, we find there is another interesting state
$p\bar{\Sigma}$, which its energy is smaller than the sum of masses of
$p$ and $\bar{\Sigma}$, but higher than the sum of masses of $p$ and
$\bar{\Lambda}$. So it may appears as a resonance state in the
$p\bar{\Lambda}$ scattering process. The calculated results are
shown in Table II and III, where $ub$ means unbound, set I (II) stands
for the calculation without (with) the annihilation interactions.
\begin{table}
\centering
\caption{The masses and widths of the state $p\bar{\Sigma}$ with
$J=0$. The theoretical threshold of $p\bar{\Sigma}$ is
2176.5 (2191.8) MeV for QDCSM (ChQM). unit: MeV}

\begin{tabular}{ccccccc}
\hline
 &\multicolumn{3}{c}{\rm QDCSM}&\multicolumn{3}{c}{\rm
 ChQM}\\ \hline
  Set~~~   & M (sc) & M (cc) & $\Gamma$ & ~~~M (sc) & M (cc) & $\Gamma$   \\ \hline
   I~~~   &  2131.8 &  2134.1   &   6.5    &   ~~~2164.8 &
   2165.6 & 7.6   \\  \hline
   II~~~   &  2174.8 & 2172.9   &   0.15    &   ~~~$ub$ &  $ub$  &  --    \\
\hline
\end{tabular}
\end{table}

\begin{table}
\centering
\caption{The same as Table 2 with $J=1$.}

\begin{tabular}{c|ccc|ccc}
\hline
 &\multicolumn{3}{|c|}{\rm QDCSM}&\multicolumn{3}{|c}{\rm
 ChQM}\\ \hline
  Set   & M (sc) & M (cc) & $\Gamma$ & M (sc) & M (cc) & $\Gamma$   \\ \hline
   I   &  2152.9 & 2151.9   &   3.1    &   2191.8 &
   2189.3 & 7.0   \\  \hline
   II   &  $ub$  & $ub$ &  --    &   $ub$ & $ub$   &  -- \\
\hline
\end{tabular}
\end{table}

\begin{figure*}[ht]
\centering
\epsfxsize=5.5in \epsfbox{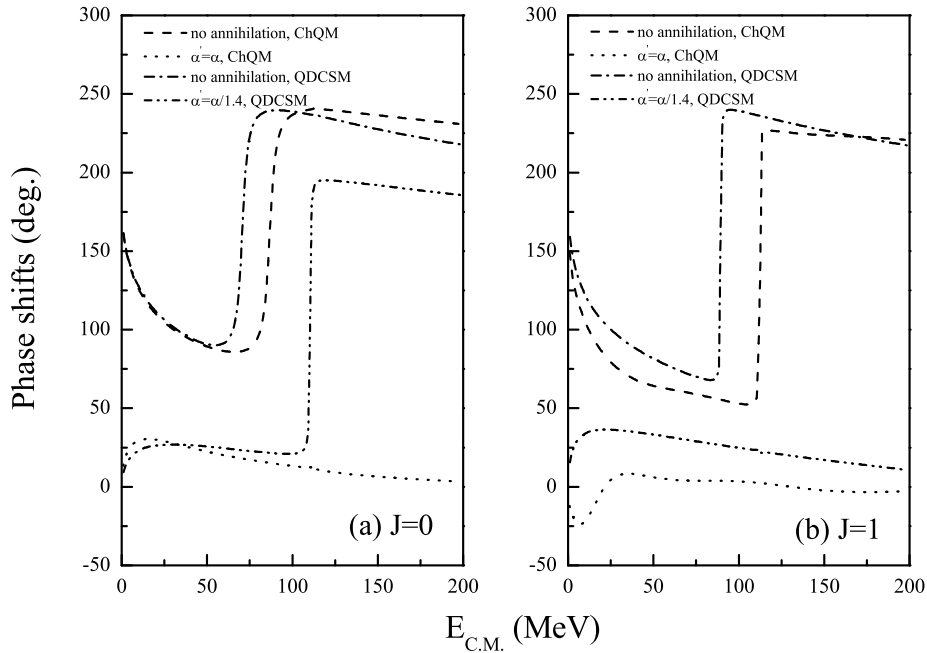}

\caption{$p\bar{\Lambda}$ $S$-wave scattering phases in ChQM and
QDCSM.}
\end{figure*}

\begin{figure*}
\centering
\epsfxsize=5.5in \epsfbox{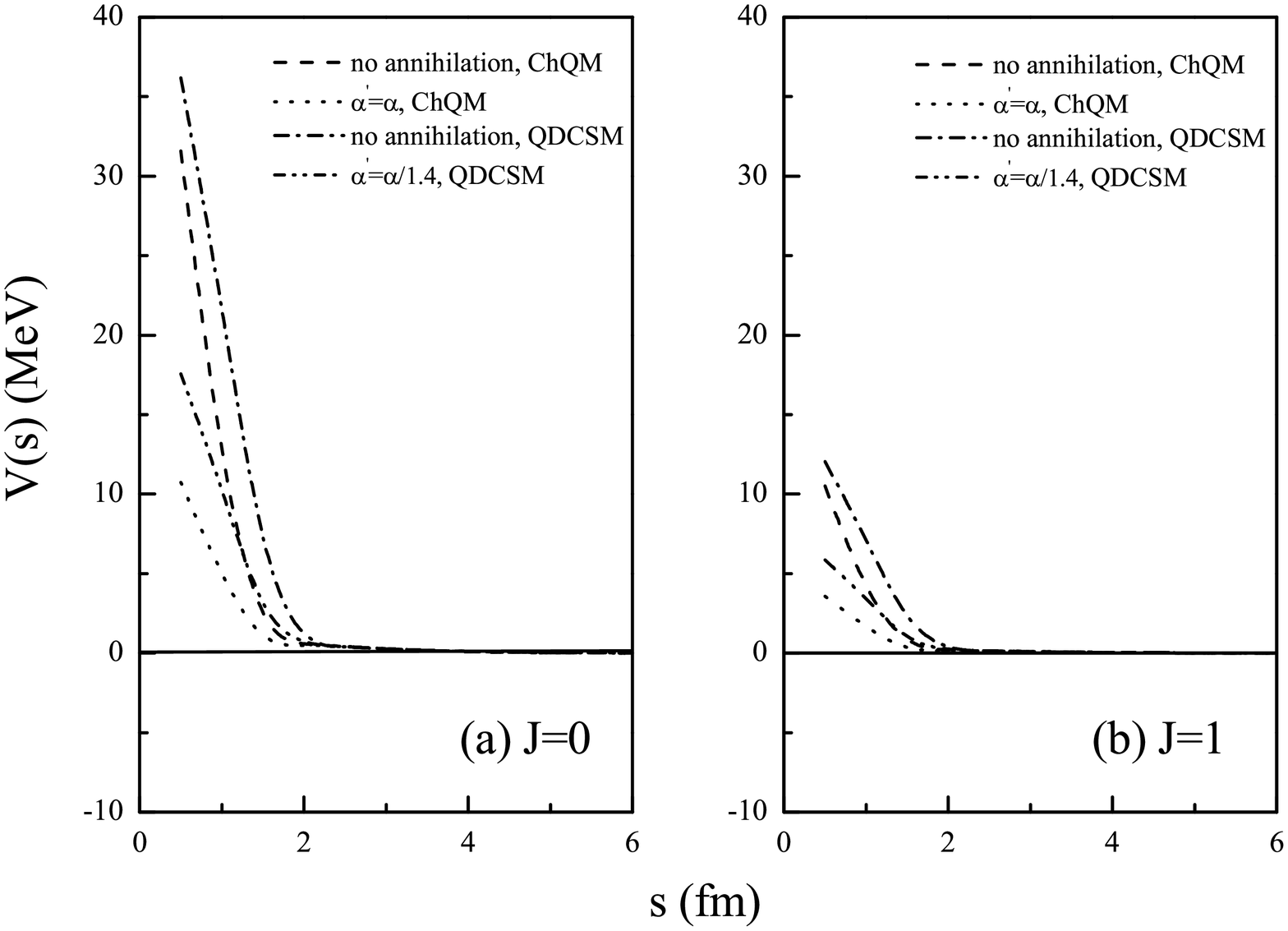}

\caption{The transition potential for $p\bar{\Lambda}-p\bar{\Sigma}$ in QDCSM.}
\end{figure*}

The single channel (sc) calculations of $p\bar{\Sigma}$ states show that
$p\bar{\Sigma}$ states are bound states if the annihilation interactions
are not included in both models. The state with spin 0
is more bound than the one with spin 1, even it is still bound with
the inclusion of the annihilation interactions in QDCSM.
After including the annihilation interactions, the state with spin 1 becomes
unbound in both models, the state with spin 0 is also unbound in ChQM, while
it is still bound state in QDCSM.
When the state $p\bar{\Sigma}$ couples to the open channel $p\bar{\Lambda}$,
the bound state will change into an elastic resonance in the
$p-\bar{\Lambda}$ scattering. The $S$-wave phase shifts of $p\bar{\Lambda}~(J=0,1)$
are calculated, and the effect of channel-coupling with $p\bar{\Sigma}$ are
taken into account. The $s$-wave phase shifts of $p\bar{\Lambda}~(J=0,1)$
are illustrated in Fig. 2. From the Fig. 2, we can see that the
phase shifts of $p\bar{\Lambda}$ rise though $\pi/2$ at
resonance masses, which listed in Tables II and III, for $J=0$ (set I and II)
and $J=1$ (set I) in QDCSM. The resonance masses are a little larger or smaller than
the energies obtained from single-channel calculations. Generally, if there
is a bound state of $p\bar{\Lambda}$, then the resonance mass of $p\bar{\Sigma}$
will be pushed up comparing with its stand alone mass, otherwise the resonance
mass will be pulled down. From the Table II, the resonance mass for set I
is pushed up, so it may infer that there is a bound state of $p\bar{\Lambda}$.
However, from the single-channel and channel coupling calculations, we do not
find a bound state for $p\bar{\Lambda}$. Maybe there is a zero-energy resonance
of $p\bar{\Lambda}$. Further study is needed. Comparing the results with
set I and II, it is clear that the annihilation interactions
play an non-negligible role in the baryon-antibaryon systems.
The inclusion of annihilation interactions pushes the state $p\bar{\Sigma}$
with spin 0 up about 40 MeV and pushes the state $p\bar{\Sigma}$ state with
spin 1 above the threshold. So from our calculation, there is a
$(p\bar{\Sigma})_{J=0}$ resonance state with resonance mass $2172.9$ MeV
and decay width $0.15$ MeV in QDCSM in the
$(p\bar{\Lambda})_{J=0}$ scattering phase shifts. The small width
comes from the fact that only $\pi$-exchange contributes and the effect of 
$\pi$-exchange is greatly reduced due to no exchange term between particle 
and antiparticle. For example, for nucleon-hyperon system, the $p\Lambda-p\Sigma$
transition potential at separation 0.5 fm is about 1800 MeV, the direct term
contributes 90 MeV only.  Fig.~3 shows the transition potential for
$p\bar{\Lambda}-p\bar{\Sigma}$. Clearly, the coupling between $p\bar{\Lambda}$
and $p\bar{\Sigma}$ is larger if the annihilation interactions are not taken into
account. Including the annihilation interactions, the transition potential will
be reduced. So the decay width in the case I is larger.
However the repulsive nature of the annihilation interaction will push
the energies of $p\bar{\Sigma}$ high in the case II, so the energy of
$p\bar{\Sigma}$ is higher in the case of II than that in case I.
Therefore we have a paradox: the state with larger binding energy has
a larger decay width. In fact, these are two different calculations,
where different interactions are used.
No $p\bar{\Sigma}$ resonance will appear in $(p\bar{\Lambda})_{J=1}$
scattering phase shifts. However, in the ChQM, no resonance can be found
if the annihilation interactions are included.

\section{SUMMARY}

In summary, we perform a dynamical study of $p\bar{\Lambda}$
systems with $J=0$ and $J=1$ in the framework of the ChQM
and the QDCSM by solving the RGM equation. All the model parameters are
taken from our previous work, which gave a good description of the
proton-antiproton $S$-wave elastic scattering cross section
experimental data. The numerical results show that the
$p\bar{\Lambda}$ systems with both $J=0$ and $J=1$ are bound
states in these two quark models if the annihilation interaction
is neglected. When the annihilation interaction is considered, the
$p\bar{\Lambda}$ systems become unbound. At the same time, the
$p\bar{\Lambda}$ elastic scattering processes with coupling to
$p\bar{\Sigma}$ state are also investigated.
The calculated phase shifts are qualitatively similar in these two
quark models. The results show that, there is no $S$-wave bound
state of $p\bar{\Lambda}$ as indicated by an enhancement near the threshold of
$p\bar{\Lambda}$ in $J/\psi$. However, it is worthy of notice that
QDCSM gives a $IJ = \frac{1}{2}0$ $p\bar{\Sigma}$
resonance state, and the state become unbound in ChQM if a single
effective scalar exchange is used in the strange system
to replace the $\sigma$ meson used in the study of $p\bar{p}$.

It is generally believed that to describe the nucleon-nucleon interaction,
the effect of one-gluon-exchange in quark model can be mimiced by the 
vector-meson exchanges in one-boson exchange model~\cite{rho}. However,
for nucleon-antihyperon conversion process, $N\bar{\Lambda}-N\bar{\Sigma}$,
one-gluon-exchange gives null contribution, while $\rho$-meson has nonzero
contribution. So this conversion process is a good place to test the
two mechanisms of baryon-antibaryon interaction in the short-range part.

Obviously our conclusion is based on the assumption that both ChQM
and QDCSM, which gave a good description of the proton-antiproton
$S$-wave elastic scattering cross section experimental data, are
suitable for $p\bar{\Lambda}$ system. In addition we assume that
the $p\bar{\Lambda}$ system is in a $(q^{3})-(\bar{q}^{3})$
configuration. More elaborating study of $p\bar{\Lambda}$ system
is worth doing in the future.

\acknowledgments{This work is supported partly by the National
Science Foundation of China under Contract Nos. 11035006, 11175088,
10947160 and the fund of Open Research Project for large
scientific instrument from the China Academy of Science.}


\begin{thebibliography}{99}
\bibitem{BES} J. Z. Bai, {\em et al.} [BES Collaboration], Phys. Rev. Lett.
{\bf 91}, 022001 (2003).
\bibitem{Belle} K. Abe, {\em et al.} [Belle collaboration], Phys. Rev. Lett.
{\bf 88}, 181803 (2002); {\bf 89}, 151802 (2002).
\bibitem{Belle2} M. Z. Wang, {\em et al.} [Belle collaboration], Phys. Rev. Lett.
{\bf 90}, 201802 (2003).
\bibitem{BES2} M. Ablikim, {\em et al.} [BES Collaboration], Phys. Rev. Lett.
{\bf 93}, 112002 (2004).
\bibitem{Belle3} Q. L. Xie, {\em et al.} [Belle collaboration], Phys. Rev. {\bf D 72},
 051105 (2005).
\bibitem{nonet} C. Z. Yuan, X. H. Mo and P. Wang, Phys. Lett. {\bf B 626},
  95 (2005).
\bibitem{Ding} G. J. Ding, J. L. Ping and M. L. Yan, Phys. Rev. {\bf D 74},
 014029 (2006).
\bibitem{Lattice} N. Ishii, S. Aoki and T. Hatsuda, Phys. Rev. Lett. {\bf 99},
 022001 (2007).
\bibitem{DSE} A. Krassnigg and C. D. Roberts, Nucl. Phys. {\bf A 737}, 7 (2004).
\bibitem{manohar}A. Manohar and H. Georgi, Nucl. Phys. {\bf B 234}, 189 (1984).
\bibitem{chiralmodel1} A. Valcarce, H. Garcilazo, F. Fernandez and P. Gonzalez,
Rep. Prog. Phys. {\bf 68}, 965 (2005) and references there in.
\bibitem{chiralmodel2} Y. Fujiwara, C. Nakamoto and Y. Suzuki,
Phys. Rev. {\bf C 54}, 2180 (1996).
\bibitem{chiralmodel3} Z. Y. Zhang, Y. W. Yu, P. N. Shen, {\em et al.},
Nucl. Phys. {\bf A 625}, 59 (1997); L. R. Dai, Z. Y. Zhang, Y. W. Yu,
and P. Wang, Nucl. Phys. {\bf A 727}, 321 (2003).
\bibitem{sigma} N. Kaiser, S. Grestendorfer and W. Weise, Nucl.
Phys. {\bf A 637}, 395 (1998); E. Oset, H. Toki, M. Mizobe and T. T. Takahashi,
Prog. Theo. Phys. {\bf 103}, 351 (2000); M. M. Kaskulov and
H. Clement, Phys. Rev. {\bf C 70}, 014002 (2004).
\bibitem{QDCSM1} F. Wang, G. H. Wu, L. J. Teng and T. Goldman,
Phys. Rev. Lett. {\bf 69}, 2901 (1992); G. H. Wu, L. J. Teng, J. L.
Ping, F. Wang and T. Goldman, Phys. Rev. {\bf C 53}, 1161 (1996); G. H. Wu,
J. L. Ping, L. J. Teng, F. Wang and T. Goldman, Nucl. Phys. {\bf A 673},
(2000) 279; J. L. Ping, F. Wang and T. Goldman, Nucl. Phys. {\bf A 657},
 95 (1999).
\bibitem{chenlz}L. Z. Chen, H. R. Pang and H. X. Huang, J. L. Ping and F. Wang, 
 Phys. Rev. {\bf C 76},  014001 (2007).
\bibitem{wong} J. L. Ping,  H. X. Huang, H. R. Pang, F. Wang and C. W. Wong,
Phys. Rev. {\bf C 79}, 024001 (2009).
\bibitem{cm} Mei Chen, H. X. Huang, J.L. Ping and F. Wang, Phys. Rev. {\bf C 83},
 015202 (2011).
\bibitem{pp} H. X. Huang, H. R. Pang and J. L. Ping, Mod. Phys. Lett. {\bf A 21}, 1231 (2011).
\bibitem{PRC73} D. R. Entem and F. Fernandez, Phys. Rev. {\bf C 73}, 045214 (2006).
\bibitem{pang} H. R. Pang, J. L. Ping and F. Wang, Chin. Phys. Lett. {\bf 25}, 3192 (2008).
\bibitem{CTP43} C. H. Chang and H. R. Pang, Commun. Theor. Phys. {\bf 43}, 275 (2005).
\bibitem{PRC62} D. R. Entem, F. Fernandez and A. Valcarce, Phys. Rev. {\bf C 62},
 034002 (2000).
\bibitem{hypertriton} H. Garcilazo, T. Fern\'{a}ndez-Caram\'{e}s and A. Valcarce,
Phys. Rev. {\bf C 75}, 034002 (2007).
\bibitem{PRD26} A. Faessler, G. L\"{u}beck and K. Shimizu, Phys. Rev. {\bf D 26},
 3280 (1982).
\bibitem{PRC65} J. L. Ping, H. R. Pang, F. Wang and T. Goldman,
Phys. Rev. {\bf C 65}, 044003 (2002).
\bibitem{rho} L. R. Dai, Z. Y. Zhang and Y. W. Yu,
 Int. J. Mod. Phys. {\bf A 20}, 1994 (2005).
\end{thebibliography}
\end{document}